\documentclass[%
reprint,
superscriptaddress,
 amsmath,amssymb,
 aps,
prb,
longbibliography,
 lengthcheck,%
]{revtex4}

\usepackage{graphicx}
\usepackage{dcolumn}
\usepackage{bm}
\usepackage{hyperref}

\begin{document}

\preprint{APS/123-QED}

\title{Magnetodielectric coupling of infrared phonons in single crystal Cu$_{2}$OSeO$_{3}$}
\author{K. H. Miller}
\affiliation{Department of Physics, University of Florida, Gainesville, Florida 32611-8440, USA}
\author{X. S. Xu}
\affiliation{Materials Science and Technology
Division, Oak Ridge National Laboratory, Oak Ridge, Tennessee 37831}
\author{H. Berger}
\affiliation{Institute of Physics of Complex Matter, Ecole Polytechnique Fe´de´ral de Lausanne, CH-1015 Lausanne, Switzerland}
\author{E. S. Knowles}
\affiliation{Department of Physics, University of Florida, Gainesville, Florida 32611-8440, USA}
\author{D. J. Arenas}
\affiliation{Department of Physics, University of Florida, Gainesville, Florida 32611-8440, USA}
\author{M. W. Meisel}
\affiliation{Department of Physics, University of Florida, Gainesville, Florida 32611-8440, USA}
\author{D. B. Tanner}
\affiliation{Department of Physics, University of Florida, Gainesville, Florida 32611-8440, USA}

\begin{abstract}
Reflection and transmission as a function of temperature have been measured on a single crystal of the magnetoelectric ferrimagnetic compound Cu$_{2}$OSeO$_{3}$ utilizing light spanning the far infrared to the visible portions of the electromagnetic spectrum. The complex dielectric function and optical properties were obtained via Kramers-Kronig analysis and by fits to a Drude-Lortentz model. The fits of the infrared phonons show a magnetodielectric effect near the transition temperature ($T_{c}\sim 60$~K).  Assignments to strong far infrared phonon modes have been made, especially those exhibiting anomalous behavior around the transition temperature.  

\end{abstract}

\maketitle

\section{Introduction}
The magnetodielectric effect refers to a change in the dielectric constant induced by an external magnetic field or by the onset of spontaneous magnetization.\cite{PhysRevB.78.094416}  Observable anomalies in the polarization of a material near a magnetic transition indicate the presence of finite magnetoelectric coupling, {\it i.e.,\/} a cross coupling between magnetic and electric orders.\cite{NatMater.7.425}  Magnetoelectric coupling has attracted significant interest\cite{NatMater.429.392,PhysCondMatt.16.9059,PhysRevB.64.054415}  recently not only for scientific purposes but also for its use in novel technological devices.  One can envision field sensors and magnetic memory switches where the magnetic order could be altered by adjusting the electric field.  A shortcoming of magnetoelectrics has been the scarcity of materials that can simultaneously support magnetic and electric orders.  The fundamental issue behind this scarcity does not lie in the symmetry demands of the two orders; rather, the particular electron configurations that favor their origins are of a contradictory nature.\cite{JPhysChem.104.6694}   The electron configurations required by magnetism usually prevail and the electric ordering is forced to originate from "improper" means.\cite{NatMater.6.13}  Recent studies on materials such as TbMn$_{2}$O$_{5},$\cite{PhysRevLett.93.177402} DyMn$_{2}$O$_{5},$\cite{PhysRevB.73.100406} and BiMnO$_{3}$\cite{PhysRevB.75.220101}  all show magnetodielectric effects near  magnetic transition temperatures where a distortion of the unit cell is also observed, as required by magnetic ordering.  This evidence suggests that the change in lattice constant, which results in a corresponding change in bond lengths, is the mechanism that alters the polarity at $T_{c}$ and, hence, produces a dielectric anomaly.

Here we report our infrared studies on a single crystal of Cu$_{2}$OSeO$_{3}$, a piezoelectric with a ferrimagnetic transition temperature of $T_{c}\sim 60$ K.\cite{PhysRevB.78.094416} Although we found no drastic anomalies across $T_{c}$, a thorough inspection of the data combined with some modeling lead us to a magnetodielectric effect.  Recently Gnezdilov {\it et al.\/}\cite{Gnezdilov} have presented a Raman study of Cu$_{2}$OSeO$_{3}$ prepared in the same way as our crystal.  They observed the abrupt appearance of 3 new lines in the spectra upon cooling below $T_{c}$, and an additional 2 lines that appeared below 20 K.   Gnezdilov {\it et al.\/} also gave a detailed description of the crystal and magnetic symmetry of this compound.  

\section{Experimental Procedures}
Single crystals of Cu$_{2}$OSeO$_{3}$ were grown by a standard chemical vapor phase method. Mixtures of high purity CuO (Alfa-Aesar, 99.995\%) and SeO$_{2}$ (Alfa-Aesar, 99.999\%)
 powder in molar ratio 2:1 were sealed in the quartz tubes with electronic grade HCl as the transport gas for the crystal growth. The ampoules were then placed horizontally into a tubular two-zone furnace and heated slowly by $50^\circ$~C/h to $600^\circ$~C. The optimum temperatures at the source and deposition zones for the growth of single crystals have been found to be $610^\circ$~C and $500^\circ$~C, respectively. After six weeks, many dark green, indeed almost black, Cu$_{2}$OSeO$_{3}$ crystals with a maximum size of $8\times6\times3$~mm$^3$ were obtained. X-ray powder diffraction (XRD) analysis was conducted on a Rigaku X-Ray diffractometer with Cu K$\alpha$ radiation ($\lambda = 1.5418$~\AA).  An electron microprobe was used for chemical analysis of all solid samples. 

Effenberger and Pertlik\cite{MonatschChem.117.887} solved the crystal structure using single-crystal X-ray diffraction.  The compact crystal structure consists of three basic building blocks, square pyramidal CuO$_{5}$, trigonal bipyramidal CuO$_{5}$, and a lone pair containing tetrahedral SeO$_{3}$ unit.  The oxygen atoms in the unit cell are shared amongst the three building blocks. The square pyramidal CuO$_{5}$ units exist in a 3-to-1 ratio to the trigonal bipyramidal CuO$_{5}$ units within the conventional unit cell.  This ratio will be important subsequently when explaining the magnetic structure.  All copper ions possess a +2 oxidation state.  More detailed descriptions of the crystal structure are found elsewhere.\cite{PhysRevB.78.094416,JLowTempPhys.39.176}

The material crystallizes in the P2$_{1}$3 cubic space group and has been shown to remain metrically cubic with no abnormal change in the lattice constant through the Curie temperature and down to 10 K.\cite{PhysRevB.78.094416}  The onset of magnetic order does have the effect of reducing the crystal symmetry to R3.  Full cubic symmetry would require all copper ions to feel the same Coulomb interaction from nearest neighbor copper spins.  The proceeding explanation is the case for ferromagnetism and antiferromagnetism but not for ferrimagnetism, which is why a reduction from cubic symmetry must accompany this ordering.  

The temperature dependent (5--300~K) reflectance and transmittance measurements were collected using a Bruker 113v Fourier Transform interferometer in conjunction with a helium cooled silicon bolometer detector in the spectral range 30--700 cm$^{-1}$ and from 700--5,000~cm$^{-1}$ with a nitrogen cooled MCT detector.  Room temperature measurements from 5,000-40,000 cm$^{-1}$ were obtained with a Zeiss microscope photometer.  After measuring the bulk reflectance over the entire spectral range, the crystal was polished to a thickness of 194 $\mu$m for transmittance measurements.  All measurements were performed using unpolarized light at near-normal incidence.  Magnetic measurements were performed in a commercial SQUID magnetometer (Quantum Design MPMS-XL7) on a single crystal sample mounted with the [111] axis parallel to the applied field.  After cooling the sample in zero field to 50~K, magnetization was measured in an applied field of 10~G while warming to 70~K.  The dc susceptibility was calculated in the low-field limit as $\chi(T) = M(T)/H$.  In addition, the isothermal magnetization as a function of applied field was measured at a temperature of 2~K, while sweeping the field from zero to 2~kG and back to zero.  
 
\section{Results and Analysis}
\subsection{Magnetism}
\begin{figure}
\includegraphics[width=3.375 in]{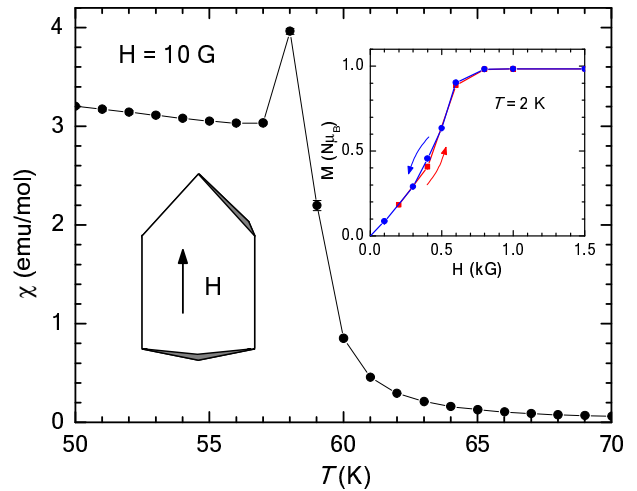}%
\caption{ The dc susceptibility, $\chi(T)$, of Cu$_{2}$OSeO$_{3}$ near the ordering temperature ($T_c$~=~60~K) in an applied field of 10~G as measured while warming after zero-field cooling to 50~K.  The inset shows the isothermal magnetization, M(H), as a function of applied field at 2~K, where the field was swept up to 2~kG before being reduced to zero.  In all instances, the lines connecting the data points are guides for the eyes. The schematic shows the orientation of the single crystal with respect to the field applied parallel to the [111] direction.}
\label{magnetism}
\end{figure}

Recent studies have measured the magnetic susceptibility of powdered samples Cu$_2$OSeO$_3$, finding ordering temperatures of $T_c^{\mathrm{inflection}}=55$~K\cite{MaterResBull.44.1} and $T_c^{\mathrm{onset}}=60$~K.\cite{PhysRevB.78.094416} Because anomalies in the infrared spectrum at the transition temperature are important, an accurate determination of $T_{c}$ for the single crystal of interest was desired.  The measured dc susceptibility as a function of temperature, $\chi(T)$, is shown in Fig.~\ref{magnetism}.  Taking the transition temperature to be where the susceptibility turns upward, $T_c^{\mathrm{onset}}=60$~K is found, consistent with the observations of Bos {\it et al.\/}\cite{PhysRevB.78.094416} At 2~K (Fig.~\ref{magnetism} inset), well within the ordered state, the magnetization saturates in a field of 800~G at 1.0~$N\mu_B$, which is half of the expected saturation value for a $S = 1/2$ system, indicating a ferrimagnetic ordering in a three-up and one-down configuration.  No coercive field was measured; however, an inflection point with some slight hysteresis was observed near 400~G, which is also consistent with the findings of Bos {\it et al.\/}\cite{PhysRevB.78.094416}

\subsection{Reflectance and Transmittance Spectra} 
\begin{figure}[htp]
\includegraphics[width=0.45\textwidth]{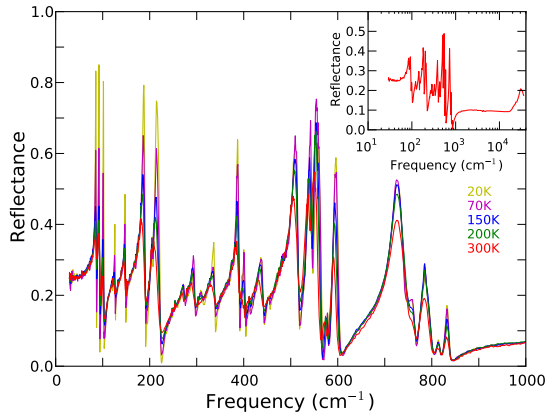}                
 \caption{(Color online) Temperature-dependent reflectance spectrum of Cu$_{2}$OSeO$_{3}$. The inset shows the 300~K reflectance out to 40,000 cm$^{-1}$.}   
\label{Refl_temps}
\end{figure}

The temperature dependent reflectance spectrum of Cu$_{2}$OSeO$_{3}$ between 30 and 1,000 cm$^{-1}$ (4--120 meV) is shown in Fig.~\ref{Refl_temps}.  A strong sharpening of many phonon modes is observed with decreasing temperature.  It should be noted that
there are no drastic anomalies in the far-infrared spectrum,
such as the presence of new modes or the splitting
of existing modes, consistent with no change in lattice constant at $T_{c}$.  Infrared spectroscopy is extremely sensitive
to changes in dipole moments, especially those
brought about by lattice distortions.  Therefore, the absence of these anomalies gives strong support to the assertion of no lattice distortions at $T_{c}$ as initially determined by X-ray diffraction measurements.\cite{PhysRevB.78.094416}  The inset of Fig.~\ref{Refl_temps} shows the 300~K reflectance up to 40,000~cm$^{-1}$.  The low absorption in the infrared is consistent with the insulating nature of the compound.  The onset of electronic absorption is indicated by the upturn of the spectra around 26,000 cm$^{-1}$ (3.2 eV). 
\begin{figure}[htp]
\includegraphics[width=0.45\textwidth]{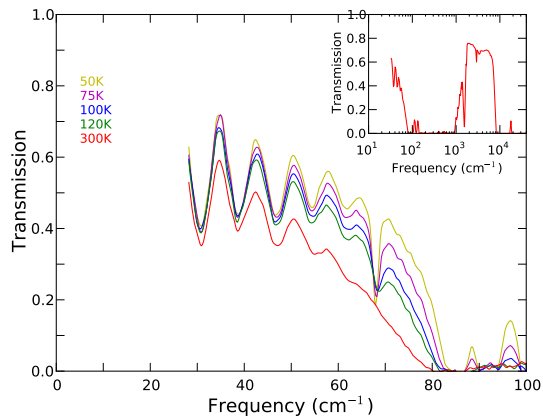}                
 \caption{(Color online) Temperature dependent transmission spectrum of 194 $\mu$m thick Cu$_{2}$OSeO$_{3}$ single crystal below 100~cm$^{-1}$.  This region is highlighted by a weak phonon ($\sim$ 68~cm$^{-1}$) that begins to appear around 120 K.  The inset shows transmission at 300 K measured out to the 40,000 cm$^{-1}$.}   
\label{Trans_temps}
\end{figure}

The transmittance spectra, as depicted in Fig.~\ref{Trans_temps}, are in good agreement with the reflectance measurements.  At frequencies below the strong phonon modes ($<80$~cm$^{-1}$), the crystal transmits.  The periodic oscillations in this range are the Fabry-Perot interference fringes due to multiple internal reflections.  Transmission gaps open between the infrared phonon modes.  These regions become increasingly more evident as temperature is lowered and the modes sharpen.  The low-frequency transmission spectrum exhibits a weak phonon with a resonance frequency of 68 cm$^{-1}$ that first appears as a small structure around 120 K and strengthens with decreasing temperature.  The mode has a small oscillator strength, thus explaining why it was not observed in reflection.  The weak phonon shows no response to a field of 0.14 kG parallel to the crystal surface; however, a magnetic origin for this structure cannot be ruled out and is worthy of future investigation. 

\subsection{Kramers-Kronig analysis and Optical Properties}

\begin{figure}[h]
\includegraphics[width=0.45\textwidth]{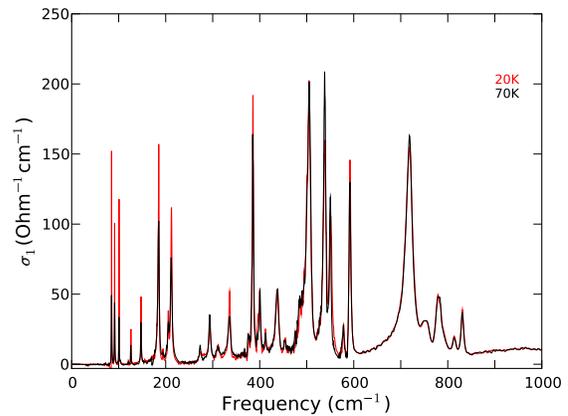}                
 \caption{(Color online) Far Infrared optical conductivity Cu$_{2}$OSeO$_{3}$ at 20 K(red) 70 K(black).}  
\label{Sigma_1}
\end{figure}

Kramers-Kronig analysis can be used to estimate the real and imaginary parts of the dielectric constant from the bulk reflectance $R(\omega$).\cite{Wooten}  
Before calculating the Kramers-Kronig integral, the low frequency data were extrapolated as a constant for $\omega \rightarrow 0$ as befits an insulator.  At high frequencies the reflectance was assumed to be constant up to $1 \times 10^7$~cm$^{-1}$ after which $R\sim(\omega)^{-4}$ was assumed as the appropriate behavior for free carrier. The optical properties were derived from the measured reflectance and the Kramers-Kronig-derived phase shift on reflection; in particular, we estimated the real part of the optical conductivity, $\sigma_{1}(\omega)$. Figure~\ref{Sigma_1} depicts $\sigma_{1}(\omega)$ at 20 K (below $T_{c}$) and 70 K (above $T_{c}$) from 30 to 1,000~cm$^{-1}$. The principal effect is a sharpening of most modes, yielding a larger conductivity at the resonant frequency. 

The phonon modes in reflectance are viewed as Lorentzian oscillators (derivative-like features) in the optical conductivity that makes them intuitive for modeling as harmonic oscillators. The vanishingly small static limit of $\sigma_{1}(\omega)$ and the low background level of conductivity throughout the infrared regions is further evidence of the insulating nature of the compound.  

\subsection{Oscillator-model fits}
\begin{figure}[h]
\includegraphics[width=0.45\textwidth]{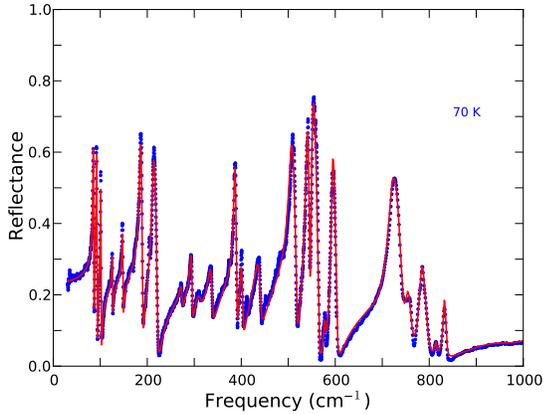}                
 \caption{(Color online) Experimental reflectance (blue points) and calculated reflectance (red line) from the Drude-Lorentz model of the Cu$_{2}$OSeO$_{3}$ 70 K dielectric function.}  
\label{Fit_refl_70K}
\end{figure}

The Drude-Lorentz model was used to fit the reflectance and obtain a second estimate of the complex dielectric function in the infrared range.  The model consists of a damped oscillator for each putative phonon in the spectrum plus a high frequency permittivity $\varepsilon_\infty$ that describes the contribution of all electronic excitations.  The model has the following mathematical form:
\begin{equation}
\varepsilon = \displaystyle\sum_{j=1}^{\infty}\frac{S_j\omega_j^2}{{{\omega_j}^2}-{{\omega}^2}-{i\omega\gamma_j}} +\varepsilon_\infty
\label{epsilonDL}
\end{equation}
where $S_j$, $\omega_j$, and $\gamma_j$ represent the oscillator strength, center frequency, and linewidth of the $j$th damped oscillator.  The complex dielectric function provided by the Drude-Lorentz model is used to calculate the reflectivity, which agrees well with our original measured quantity.  Figure~\ref{Fit_refl_70K} shows the calculated reflectivity and the measured reflectivity at 70 K. 
 \begin{figure}[htp]
\includegraphics[width=0.45\textwidth]{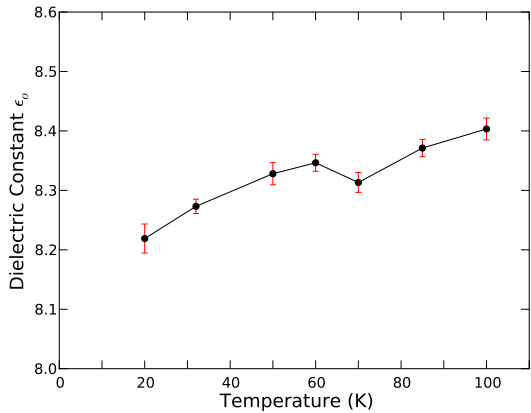}                
 \caption{The static dielectric constant as calculated from the Drude-Lorentz model at temperatures between 20 and 100 K, including $T_{c}$.}
\label{dielect_const}
\end{figure}

 \section{Discussion}
\begin{figure*}[t]
   \centering
\includegraphics[width=0.75\textwidth]{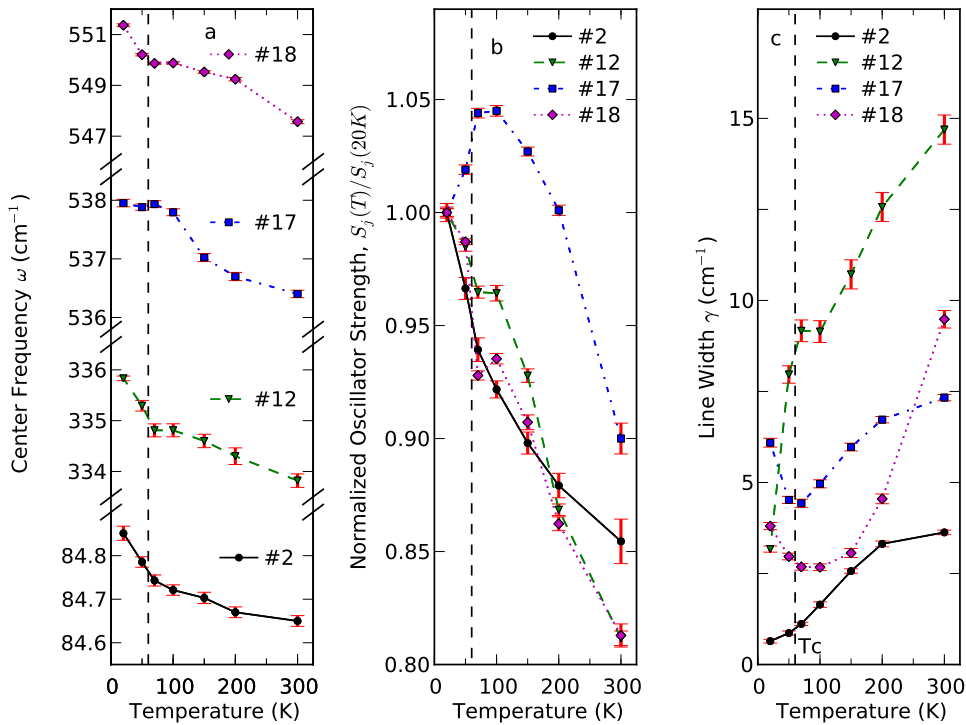}
   \caption{(Color online) The (a) center frequency, (b) normalized oscillator strength, and (c) linewidth of four oscillators as a function of temperature.  Oscillator \#3 is a typical conventional phonon, whereas the other three oscillators show anomalous behavior as temperature is lowered across $T_{c}$.}
   \label{osc_parms}
\end{figure*}

\subsection{Magnetodielectric effect}
Equipped with oscillator parameters to describe each of the infrared phonons at all measured temperatures, one is now in position to closely monitor the subtle dynamics of the phonon structures across the transition temperature.  Despite the lack of drastic changes in the phonon spectrum at $T_{c}$, it is worthwhile examining whether a combination of many small anomalies in the phonon dynamics might sum to give an overall effect.  At this point it is logical to examine the static dielectric constant because it is the sum of parameters that describe the dielectric nature of the compound (the side to which the infrared is most sensitive).  Taking the zero frequency limit of the Drude-Lorentz formula, one arrives at the following simple formula for the static dielectric constant:
\begin{equation}
\varepsilon_o = \displaystyle\sum_{j=1}^{\infty}{S_j +\varepsilon_\infty}
\label{staticepsilonDL}
\end{equation}
The calculated static dielectric constant at temperatures surrounding $T_{c}$, is shown in Fig.~\ref{dielect_const}.  There is an anomalous jump at around 60 K.  It should be noted that this magnetodielectric effect in Cu$_{2}$OSeO$_{3}$ was previously observed by Bos {\it et al.\/}\cite{PhysRevB.78.094416} through dielectric capacitance measurements.  While the value of the dielectric constant measured here is considerably smaller than that in the previous report, the direction and magnitude of the anomaly at $T_{c}$ are in good agreement. 

\subsection{Anomalous phonons}
The observed phonons in Cu$_{2}$OSeO$_{3}$ may be divided into two categories: conventional and anomalous. Conventional phonons show a slight hardening of their frequencies when cooled to low temperatures, accompanied by a significant reduction in linewidth and at most modest changes in oscillator strength. Moreover, the temperature variation is smooth with no sudden changes in slope or value. Anomalous phonons violate one or more of these expectations. We have identified 11 anomalous phonons in our spectra, which show 28 total modes. A numbering scheme has been used which corresponds to the sequence of appearance of oscillators in the infrared spectrum starting with \#1 (lowest frequency mode) and ending with \#28 (highest frequency mode)  

Figure~\ref{osc_parms} displays the temperature dependence of the three oscillator parameters for 3 of the 11 anomalous phonons in the infrared spectrum.  The oscillator parameters for one conventional phonon are shown for comparison.   It was observed that anomalous behavior usually could be found in all three oscillator parameters.  

\subsection{Assignment of phonon modes}
Since the magnetodielectric effect is observed through lattice dynamics, analysis of the phonon spectrum is incumbent. The number of phonon modes expected in Cu$_{2}$OSeO$_{3}$ can be estimated by space group analysis.  Using the SMODES program,\cite{SMODES} we arrive at the following distribution of modes:
\begin{equation}
\Gamma^{optical} = 14A^{\it(R)}+14E^{*\it(R)}+41T^{\it(R,IR)}
\label{Gamma_optical}
\end{equation}
where (R) and (IR) denote Raman active and infrared active modes respectively. We therefore have the potential of 41 total infrared active modes, all of which possess threefold degeneracy as indicated by their irreducible representation.  However, only 28 modes in the infrared spectrum are detected, and a discussion about this discrepancy between the observed and predicted modes is beyond the scope of this work.
  
We performed lattice dynamical calculations based on a real-space summation of screened coulomb interactions involving a spherical cut-off boundary.~\cite{Wolf}  Frequency, mode intensity, as well as displacement pattern were calculated based on the structure and valence as reported by Bos {\it et al.\/}~\cite{PhysRevB.78.094416}  Seven strong modes in the infrared spectrum have been assigned by comparing the calculated and experimental spectra.

Table~\ref{Table} details the mode assignments made using the adopted numbering scheme and center frequency of each oscillator for identification.  It should be noted that oscillators \#12 and \#15, which exhibit anomalous behavior at the transition temperature (\#12 is shown in Fig.~\ref{osc_parms}), are associated with vibrations of oxygen around the central copper, the ion responsible for magnetic ordering.
\begin{table*}[htp]
\caption{Oscillator parameters (at 20 K) and their corresponding assignments for a few strong modes in the far infrared.}
\begin{ruledtabular}
\begin{tabular}{l*{6}{c}r}
 &  & 20 K & & Cu$_{2}$OSeO$_{3}$ & \\
\hline
Index & Osc Strength & Center Freq & Linewidth & Anomalous & Assignment \\
 & S & $\omega$ (cm$^{-1}$) & $\gamma$ (cm$^{-1}$) &  &  \\
\hline
  8 & 0.521 & 212 & 2.9 & No & SeO$_{3}$ vibrating against CuO$_{5}$   \\
  12 & 0.107 & 336 & 3.6 & Yes & CuO$_{5}$ in plane vibration    \\
  13 & 0.188 & 385 & 2.0 & No & SeO$_{3}$ bending mode \\
  15 & 0.095 & 437 & 5.9 & Yes & CuO$_{5}$ out of plane vibration \\
  21 & 0.291 & 717 & 15.8 & No & SeO$_{3}$ antisymmetric stretch \\
  23 & 0.054 & 781 & 13.9 & No & SeO$_{3}$ antisymmetric stretch \\
  25 & 0.016 & 831 & 5.1 & No & SeO$_{3}$ radial breathing mode \\
\end{tabular}
\end{ruledtabular}
\label{Table}
\end{table*} 

\section{Conclusions}
Far-infrared measurements of single crystal reflectance from Cu$_{2}$OSeO$_{3}$ reveal no drastic anomalies in the phonon spectrum as temperature traverses $T_{c}$.  However, a closer inspection of the dynamics of the phonon spectrum, as modeled through Drude-Lorentz fitting, uncover an anomalous jump in the dielectric constant near $T_{c}$.  It is observed that 11 of the 28 total far-infrared phonons contribute to this magnetodielectric effect.  A few strong far-infrared phonons have been assigned to motion of the CuO$_{5}$ and SeO$_{3}$ units via a lattice dynamical calculation.  It is noteworthy that 2 of the 11 modes exhibiting anomalous behavior across $T_{c}$ have been assigned to the motions of oxygen around the central copper, the ion responsible for magnetic ordering.  A weak phonon that was not resolved in reflectance is observed below 120 K in the transmission spectrum.  A magnetic origin for this structure has yet to be ruled out.  
Our infrared results agree with the Raman studies of Gnezdilov {\it et al.\/}\cite{Gnezdilov} in that we also observed a number of phonon modes that exhibit anomalies in their strength, center frequency, and linewidth, but we differ on other issues.  For example, we do not observe any new modes below the magnetic transition whereas Genzdilov {\it et al.\/}\cite{Gnezdilov} do detect some additional (rather weak and broad) features.  They report three new modes appearing in the Raman spectra below the transition temperature at frequencies of $\sim$261, 270, and 420~cm$^{-1}$.  The only structure we observe in these three spectral regions is at 270~cm$^{-1}$, where 1 of the 11 reported anomalous phonons is present.  Genzdilov {\it et al.\/}\cite{Gnezdilov} also report two new lines originating below 20 K in the Raman spectra at $\sim$86 and 203~cm$^{-1}$.  Our infrared studies reveal a strong rather typical mode at $\sim$86~cm$^{-1}$ and a weak anomalous mode at 203~cm$^{-1}$.  If any new infrared features possessed the same relative intensities as reported for the Raman spectra, we would have observed them clearly.

\begin{acknowledgments}
The authors wish to thank H. T. Stokes and F. Pfuner for valuable discussions on the SMODES program.  This work was supported by DOE through grant DE-FG02-02ER45984 and the NSF via DMR-0701400.                    

\end{acknowledgments}

\end{document}